\documentstyle[aps,epsfig,twocolumn]{revtex}

\draft
\begin{document}

\twocolumn[\hsize\textwidth\columnwidth\hsize\csname
@twocolumnfalse\endcsname

\title{Rapid algorithm for identifying backbones in the
two-dimensional percolation model}
\author{Wei-Guo Yin\cite{yin}}
\address{Department of Physics, Fudan University, Shanghai 200433, People's
Republic of China\\%
and Physics Department, the Chinese University of Hong Kong, Hong
Kong, People's Republic of China}
\author{Ruibao Tao}
\address{CCAST (World Laboratory), P. O. Box 8730 , Beijing 100080, People's
Republic of China\\%
and Department of Physics, Fudan University, Shanghai 200433,
People's Republic of China}
\date{June 16, 1998; v2, September 22, 1999; v3, \today} \maketitle

\begin{abstract}
We present a rapid algorithm for identifying the current-carrying
backbone in the percolation model. It applies to
general two-dimensional graphs with open boundary conditions.
Complemented by the modified Hoshen-Kopelman cluster labeling algorithm,
our algorithm identifies dangling parts using their local properties.
For planar graphs, it finds the backbone 
almost four times as fast as Tarjan's depth-first-search
algorithm, and uses the memory of the same size as the modified
Hoshen-Kopelman algorithm. 
Comparison with other algorithms for backbone identification is addressed. 
%The scaling exponent $q/\nu$ ($\beta/\nu$) of the
%probability that a site belongs to the backbone (the spanning
%cluster) is found to be $0.3647 \pm 0.0039$ ($0.1068 \pm 0.0013$)
%and satisfy the Chayes' exponent inequality: $ 2 \beta \le q $.
\end{abstract}

%64.60.Ak   Renormalization-group, fractal, percolation studies of phase transitions
%02.70.-c   Computational techniques
%05.10.-a   Computational methods in statistical physics and nonlinear dynamics

]

The percolation model describes a system consisting of randomly
distributed ``conducting'' cells and ``isolating'' cells \cite{Stauffer}.
When the density of the conducting cells exceeds
a threshold value $p_{\rm c}$, a cluster of connected conducting cells
spans the system. This model has become one of the most
extensively studied statistical models because of its simplicity
and its often surprising applicability.
%For example, it has functioned in the studies of transportation,
%communication, and various diffusion processes of forest fires, pollution,
%epidemics, etc. \cite{Grimmett}.
Its relevance to physics includes, for example,
%the wetting phase transition and
the metal-insulator transition observed in disordered
systems \cite{Kirkpatrick,Essam}. Numerical simulation is an
important means in the solution of the model. For a complex
system, which can be originated from the complexity of
its geometry or the mechanism underlying the random distribution function,
numerical simulation seems to be a unique method available.
To get rid off severe statistical fluctuations brought by disorder
in the percolation model, numerical calculations, in general,
need to be done over a large number of large scale ensembles.
Finding rapid algorithms is thus an interesting topic in this field
\cite{Hoshen,Rintoul,Herrmann,Tarjan,Moukarzel,Ziff0,Grassberger,Lobb,%
Grassberger99,Newman,Edwards,Bastiaansen,Yin}.

The critical behavior of percolation is well understood.
Because of its relationship to the one-state Potts model
\cite{Stauffer}, all critical exponents that have thermal analogue
are exactly known for the dimension $d=2$. Those having no thermal
analogue are usually estimated by numerical work.
The most important of these are the backbone fractal dimension and
the conductivity exponent.
The backbone is defined as the subset of the spanning
cluster that carries current when a potential difference is
applied between two sides of the system.
Far away from $p_{\rm
c}$, the conductivity of the system can be accurately calculated
within the effective-medium theory \cite{Kirkpatrick,Yin2};
otherwise, numerical studies are needed. The calculation of
the conductivity on the current-carrying backbone is considerably
faster than on the spanning cluster \cite{Grassberger99,Edwards} because
the density of the backbone in the spanning cluster substantially
decreases as the system approaches to $p_{\rm c}$ \cite{Chayes}.
Backbone identification is thus a necessary step before estimating
the conductivity using the Lobb-Frank algorithm \cite{Lobb} or the
multigrid algorithm \cite{Edwards,Bastiaansen}.

The purpose of this paper is to present a fast algorithm for $d=2$
backbone identification. The backbone is called the set of
biconnected nodes in computer science. For general graphs,
heretofore the fastest algorithm to identify it is Tarjan's
depth-first-search algorithm \cite{Tarjan} which runs in time
$O(N)$ for a graph of $N$ nodes, provided the number of edges
meeting at each node is finite. For $d=2$ planar graphs, our
algorithm is {\em almost four times} as fast as Tarjan's
algorithm. We emphasize here that our algorithm applies to general
$d=2$ graphs with open boundary conditions. There are other
algorithms for backbone identification
\cite{Rintoul,Herrmann,Moukarzel,Grassberger,Grassberger99}. We
will first describe our algorithm and then compare it with these
known algorithms.

\begin{figure}[tbp]
  \begin{center}
  \epsfig{file=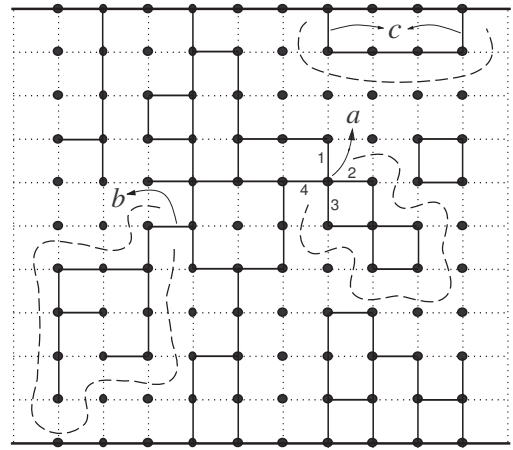,width=0.9\hsize,angle=0}
  \end{center}
%  \vskip -2cm
\caption{An illustration of percolation on a square lattice of
linear size $L=10$: site (filled circle), open bond (solid
line), closed bond (dotted line), plaquette (smallest square area). $a$
is a dangling-loop-connecting site, $b$ is an open bond connecting
a dangling end, and $c$ are open bonds belonging to a dangling
arc. Dashed lines are some paths of connected plaquettes enclosing
some dangling parts.} \label{model}
\end{figure}

Below we give the definitions of all terminologies involved
and illustrate them in a square lattice having $L$ sites horizontally
and $L+1$ sites vertically (see Figure~\ref{model}).
Between neighboring sites we insert a bond that is taken to be {\it open}
with an independent probability $p$, or {\it closed} with
probability $1-p$.
Obviously, here is defined the percolation model for bonds \cite{site}.
Open lateral boundary conditions are assumed, whereas
all bonds locating at the top and the bottom are set open.
A potential difference is applied between the top and the bottom.
Two neighboring sites are {\em connected} to each other
if the bond between them is open.
The {\it spanning cluster} is the set of sites if there is a path of
open bonds from each of them to the bottom and to the top.
The spanning cluster consists of the current-carrying backbone and
non-current-carrying dangling parts.
There are three kinds of dangling parts of the spanning cluster:
dangling ends, dangling arcs, and dangling loops.
A {\it dangling end} is connected to the spanning cluster by only
one open bond (e.g., $b$ in Figure~\ref{model});
A {\it dangling arc} disconnects with the top (bottom)
but is connected to the bottom (top) by at least two open bonds
(e.g., $c$'s in Figure~\ref{model});
A {\it dangling loop} is connected by at least two open bonds
to only one site of the spanning cluster (e.g., $a$ in Figure~\ref{model}).
Here we name this site a {\it dangling-loop-connecting site},
which is called an articulation node in computer science.
After all dangling parts are removed, the remainder of the spanning
cluster is the {\it backbone}.

Our algorithm is based on {\bf Jordan's curve theorem}: Suppose
there is a loop in a two-dimensional graph, then the area inside
the loop disconnects with the area outside, and vise versa.

To make use of the theorem, we introduce the concept of {\it
plaquette} that is an as small as possible area enclosed by a loop
of bonds which can be either open or closed. For example, any one
of the smallest square areas in Figure~\ref{model}. We define that
{\em two nearest neighboring plaquettes are connected if the bond
between them is closed}. Note that the definition of connectedness
for plaquettes is opposite to that for sites. Jordan's curve
theorem tells us that any dangling part must be enclosed by a loop
of connected plaquettes if it is split from the spanning cluster.
Hence, by use of the connectedness of plaquettes, we are able to
identify dangling parts according to the following three
corollaries of the theorem, which describe the {\em local}
properties of the three kinds of dangling parts, respectively.
We will directly use them to design our algorithm.

{\bf Corollary 1.} An open bond connects a {\em dangling end} to
the spanning cluster if and only if two plaquettes beside the bond
are connected. For example, $b$ in Figure~\ref{model}.

{\bf Corollary 2.} If two non-neighboring plaquettes nearest neighboring
the top (bottom) are connected, all sites that are connected to the
top (bottom) by the bonds between the two plaquettes belong to
{\em dangling arcs}. For example, $c$ in Figure~\ref{model}.

{\bf Corollary 3.} Let all dangling ends and arcs have been
removed. Site $a$ is a {\em dangling-loop-connecting site} if and
only if (i) $a$ connects at least four open bonds, referred to as
bonds 1,2,3,4, and (ii) among its neighboring plaquettes, some
between two of the four bonds (e.g., bonds 1 and 2) is connected
by a path of plaquettes to some others between the other two bonds
(i.e., bonds 3 and 4). For example, $a$ in Figure~\ref{model}. $a$'s
northern, eastern, southern, and western open bonds correspond to
bonds 1, 2, 3, 4, respectively. $a$'s northeast plaquette which is
between bonds 1 and 2 is connected by a path of plaquettes to
$a$'s southwest plaquette which is between bonds 3 and 4. Note
that here bonds 1 and 4 belong to the backbone, bonds 2 and 3
belong to a dangling loop.

\begin{figure}[tbp]
  \begin{center}
  \epsfig{file=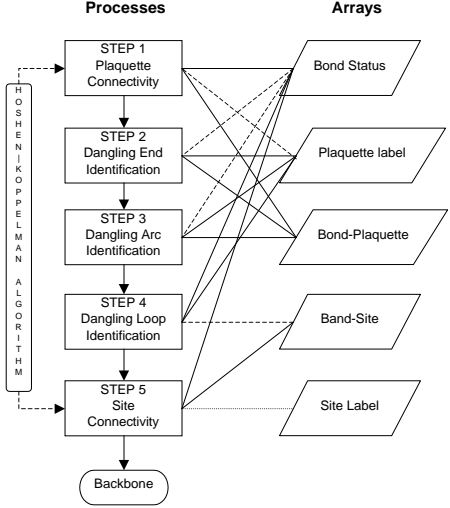, width=1.2\hsize,angle=0}
  \end{center}
  \vskip -4.8 cm
\caption{Flowchart of our algorithm. A line between an array
(parallelogram) and a process (rectangle) indicates that the array
is used in the process. If the line is dashed, the array changes
in the process.} \label{memory}
\end{figure}

Now let us describe the procedure of our algorithm. 
In accordance with open lateral boundary conditions, 
the left (right) plaquettes
of the leftmost (rightmost) column of vertical bonds are set to be
the same, respectively, as shown in Figure~\ref{model}. 
The algorithm is carried out as follows (see
Figure~\ref{memory}).

{\bf Step 1.} Compute the connectedness of plaquettes. 
If the leftmost plaquette is connected to the rightmost plaquette 
by a path of plaquettes, the spanning cluster does not exist
\cite{note}. 

{\bf Step 2} for identifying dangling ends. Sweep the bond graph
to close any bond whose two-side plaquettes have the same label.
This is the application of Corollary 1.

{\bf Step 3} for identifying dangling arcs. If two plaquettes
neighboring the top (or the bottom) have the same label, then
close all bonds between the two plaquettes. This is the
application of Corollary 2.

{\bf Step 4} for identifying dangling loops. We look for
dangling-loop-connecting sites ($a$) using Corollary 3. That is,
$a$ connects four open bonds, referred to as bonds 1,2,3,4, and a
plaquette between bond 1 and 2 has the same label as a plaquette
between bond 3 and 4. Note that $a$'s themselves belong to the
backbone. We use the following trick to split dangling loops: We
create a new site $a^\prime$ and let bonds 2 and 3 be connected to
$a^\prime$ instead of $a$. As a result, the subgraph connected to
bonds 2 and 3 is disconnected to that connected to bonds 1 and 4.

{\bf Step 5.} Compute the connectedness of sites including those
newly created sites ($a^\prime$). The resultant spanning cluster
is exactly the backbone.

In this paper, the connectedness of plaquettes or sites is calculated 
by using the modified Hoshen-Kopelman cluster-labeling algorithm
\cite{Hoshen,Edwards}.
The data structure of this algorithm includes a {\it bond status array} that
records the status of any bond \cite{data}, a {\it site label
array} that records the label of any site (connected sites will
have the same label), and a {\it bond-site array} that records two
end sites of any bond. Our algorithm needs two extra arrays to
record the information of plaquettes. One is a {\it plaquette
label array} that records the label of any plaquette (connected
plaquettes will have the same label), the other is a {\it
bond-plaquette array} that records two side plaquettes of any
bond. Given this data structure, we can compute both the
connectedness of sites and the connectedness of plaquettes.
The processes of the algorithm and the usage of the arrays are
summarized in Figure~\ref{memory}. Only the bond status array is
used in all steps. It is clear that if the memory is the
bottle-neck of calculations, the bond-site array and the
bond-plaquette array can occupy the same memory, so can the site
label array and the plaquette array. The reasons are (i) they are
not used simultaneously and (ii) they are used sequentially. The
bond-plaquette array is used in Step 1-3 while the bond-site array
is used in Step 4-5. The plaquette label array is used in Step 1-4
while the site label array is used in Step 5.

\begin{figure}[tbp]
  \begin{center}
  \epsfig{file=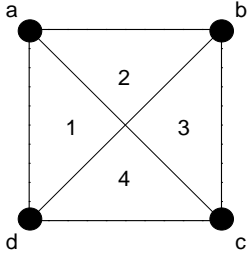,width=\hsize,angle=0}
  \end{center}
  \vskip -2.5 cm
\caption{A square lattice of linear size $L=2$ with crossing
bonds ($\overline{ac}$ and $\overline{bd}$). Filled circles
$a,b,c,d$ denote {sites}. The smallest trianglular areas $1,2,3,4$
denote {\em plaquettes}. Bond $\overline{ac}$ has two pairs of side plaquettes
[$(1,2)$ and $(3,4)$]. Bond $\overline{bd}$ has two pairs of side
plaquettes [$(1,4)$ and $(2,3)$]. If any pair of side plaquettes
of a crossing bond have the same label, the crossing bond can be
closed.} \label{nonplanar}
\end{figure}

Figure~\ref{nonplanar} shows one of the simplest $d=2$ {\em
nonplanar} graphs: a square lattice with both nearest neighboring
({\em nn}) bonds and next nearest neighboring ({\em nnn}) bonds.
This type of nonplanar graphs is important because it is often
used to mimic the effects of higher dimensionality
\cite{Bastiaansen}.
Here a {\em plaquette} is a smallest triangular area.
While a {\em nn} bond still has two side
plaquettes, a {\em nnn} bond has four which are divided into two
pairs by another {\em nnn} bond. If any pair of side plaquettes of
a {\em nnn} bond have the same label, the {\em nnn} bond can be
closed in Step 2. The number of bonds (plaquettes) in this case is
twice (four times) as large as that in the corresponding planar
square lattice with {\em nn} bonds only.

\begin{figure}[tbp]
  \begin{center}
  \epsfig{file=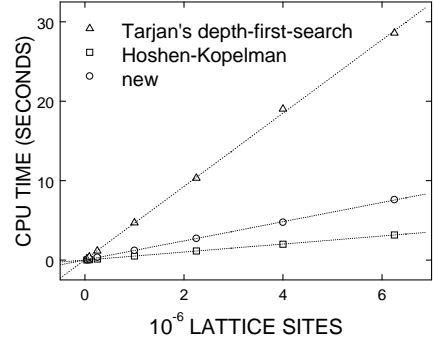,width=\hsize,angle=0}
  \end{center}
  \vskip -5. cm
\caption{CPU time of three algorithms for bond percolation at
$p=p^{\rm (b)}_{\rm c}=0.5$. The dotted lines are our fits to the
data. Their slopes are 0.51, 1.21, and 4.62 for the modified
Hoshen-Kopelman algorithm, the new algorithm, and Tarjan's
depth-first-search algorithm, respectively.} \label{cputime}
\end{figure}

The efficiency of our algorithm is demonstrated in
Figure~\ref{cputime} for square lattices with {\em nn} bonds only,
where the central-processing-unit (CPU) time needed for the
simulation is plotted versus the number of sites. We consider bond
percolation at $p=p^{\rm (b)}_{\rm c}=0.5$~\cite{Essam} and site
percolation~\cite{site} at $p=p^{\rm (s)}_{\rm c}=0.592745
(2)$~\cite{Ziff}. The results are obtained by averaging over 1000
configurations~\cite{data} for linear size
$L=100, 200, 300, 500, 1000, 1500, 2000, 2500$, respectively.
Results for systems of larger size and more configurations will be
reported soon.
The calculations are carried out on a
Pentium II/233 processor with Red Hat Linux release 4.2 operating
system and GNU project F77 Compiler (v0.5.18). As shown in
Figure~\ref{cputime}, the dotted lines are our fits to the data
for three algorithms. Their slopes are 0.51, 1.21, and 4.62 for
the modified Hoshen-Kopelman algorithm, our new algorithm, and
Tarjan's depth-first-search algorithm, respectively. All the three
algorithms run in time proportional to the system size. The total
time needed to find the backbone is about twice as many as that to
compute the connected cluster, since we use the modified
Hoshen-Kopelman algorithm twice in the former and once in the
latter. The present algorithm finds the current-carrying backbone
almost four times as fast as Tarjan's depth-first-search
algorithm. Note that in our algorithm finding the backbone is not
subject to finding the spanning cluster first. An advantage of the
new algorithm is that it is easy to program and maintain because
it involves only three local properties of dangling parts and uses the
well-known modified Hoshen-Kopelman algorithm.
The codes of the algorithm are available from us.

There are other algorithms for backbone identification
\cite{Rintoul,Herrmann,Moukarzel,Grassberger,Yin}.
A brief review of them was recently given in Ref. \cite{Grassberger99}.
Here we make a comparison between ours and these known algorithm.
The traditionally used algorithm by
physicists is the burning algorithm \cite{Rintoul,Herrmann}, which
is at least for large $N$ much slower than Tarjan's algorithm
\cite{Grassberger99}.
The matching algorithm with complexity slight larger than
$N$ [$O(N^{1.07})$ for $d=2$] was used recently in literature \cite{Moukarzel}.
For strictly planar graphs, the hull-generating algorithm is even faster
\cite{Ziff0,Grassberger}. This algorithm has the same
asymptotic complexity as Tarjan's algorithm, but it is roughly
twice as fast and uses about half of the memory, since it needs
one data structure less and needs only one pass through all sites,
instead of two passes in Tarjan's algorithm \cite{Grassberger99}.
For planar graphs, our algorithm is almost four times as
fast as Tarjan's algorithm and uses the memory of the same size as
the modified Hoshen-Kopelman algorithm.
For site percolation \cite{site}, a modified version of our algorithm
uses the memory of even smaller size where the bond-site array and
the bond-plaquette array are discarded \cite{Yin}.
Since the hull-generating algorithm, Tarjan's, and ours 
have the same asymptotic complexity, their difference in speed 
is determined by the complexity of inner operations.
Note that the identification of the backbone or the spanning cluster involves
the {\em global} properties of sites.
Our algorithm has an advantage such that
it is able to identify dangling parts using their {\em local}
properties complemented by the most efficient algorithm computing  
the connectedness of sites, which is currently  
the modified Hoshen-Kopelman algorithm.
The modified Hoshen-Kopelman algorithm was originally used to identify
the spanning cluster \cite{Hoshen}.
Because of its simplicity, the geometrical properties of the
spanning cluster have been numerically studied on very large
systems containing as much as $10^{11}$ sites \cite{Hoshen}.
Therefore, the inner operations of our algorithm are simpler than
those of Tarjan's and hull-generating.
It should be made clear that
the algorithms of Tarjan, matching, and burning can be used for
arbitrary graphs; the hull-generating algorithm is valid for strictly
planar graphs only; ours applies to general $d=2$ graphs with open
literal boundary conditions, and thus has a wider range of
application than the hull-generating algorithm.

The exact finding of the spanning cluster and the backbone
enables us to calculate some scaling exponents for percolation.
The order parameter of percolation is $P_\infty$, the ratio of
the sites in the spanning cluster to all sites in the system.
Correspondingly, we define $Q_\infty$, the ratio of the sites
in the backbone to all sites in the system.
These quantities scale with linear size $L$ as

\begin{equation}
P_\infty(L) \sim L^{-\beta/\nu}, \hspace{0.25cm} Q_\infty(L) \sim
L^{-q/\nu}, \hspace{0.25cm} {\rm for} \,\, L \rightarrow \infty
\end{equation}
where $\beta$, $q$, and $\nu$ are the critical exponents of $P_\infty$,
$Q_\infty$, and the correlation length, respectively.

\begin{figure}[tbp]
  \begin{center}
  \epsfig{file=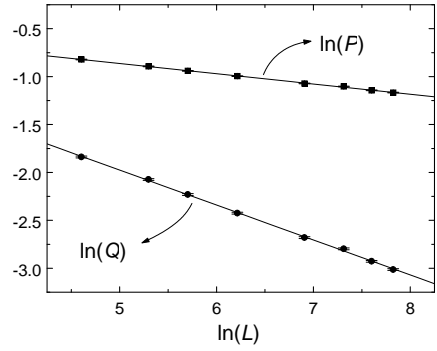, width=\hsize,angle=0}
  \end{center}
  \vskip -5 cm
\caption{Log-Log plot of $Q_\infty$ and $P_\infty$ for site
percolation at $p=p^{\rm (s)}_{\rm c}=0.5927$. The dotted lines
are the results of our weighted-least-squares fits against the
functions $Q_\infty(L)=aL^{-q/\nu}$ and
$P_\infty=bL^{-\beta/\nu}$, giving values of $q/\nu=0.3647 \pm
0.0039$ and $\beta/\nu=0.1068 \pm 0.0013$.} \label{exponent}
\end{figure}

We calculate these scaling exponents in our Monte Carlo runs.
Note that the following numerics are listed for demonstration because
we have not calculated them on a world-record-breaking number of
world-record-breaking scale lattices. The largest size
considered here is $L=2500$ with 1000 configurations.
Results for $L_{\rm max}=4096$ with 198470 configurations were reported in
Ref. \cite{Grassberger99}.
Calculations for systems of larger size and more configurations are in progress.
The data are plotted in
Figure~\ref{exponent} with very small standard errors of the mean.
The lines shown in the log-log plot are the results of the
weighted-least-squares fits, and yield the value $\beta/\nu=0.1068
\pm 0.0013$ (close to the exact result $\beta/\nu=5/48$
\cite{Stauffer,Hoshen}), and $q/\nu = 0.3647 \pm 0.0039$. The
calculated exponents satisfy the Chayes' exponent inequalities
\cite{Chayes}:

\begin{equation}
2\beta \le q \le t,
\end{equation}
where $t$ is the exponent of conductivity (the best estimate for
two-dimensional systems known to us is $t/\nu = 0.9745 \pm
0.0015$~\cite{Normand}). The critical behavior of the backbone is
consistently closer to that of the conductivity than that of the
spanning cluster. Since $Q_\infty / P_\infty $ vanishes as $p \to
p_{\rm c}$, the unknowns in the calculation of conductivity on the
backbone is considerably less than those on the spanning cluster.
As a result, the calculation of electronic conductivity on the
backbone has considerably less statistical fluctuation and much
reduced critical slowing down than on the spanning cluster
\cite{Edwards}.

Summarizing, we present a fast algorithm for identifying
percolation backbones in general two-dimensional graphs with open
boundary conditions. Complemented by the modified Hoshen-Kopelman 
algorithm, our algorithm identifies dangling parts using
their local properties. For planar graphs, our algorithm is by far the
fastest algorithm for backbone identification: it finds the
backbone almost four times as fast as Tarjan's depth-first-search
algorithm, and uses the memory of the same size as the
modified Hoshen-Kopelman algorithm.

We thank Paul Bastiaansen for providing us with the code of
Tarjan's depth-first-search algorithm implemented by J. Goodman 
(Ref. \cite{Edwards}) 
and M. E. J. Newman for pointing out 
the original hull-generating algorithm (Ref. \cite{Ziff0}) and his recent
work (Ref. \cite{Newman}). We are grateful to our
referees for their useful suggestions and 
pointing out recent important work done on the
problem of percolation backbones
(Ref. \cite{Rintoul,Herrmann,Moukarzel,Grassberger,Grassberger99}). 
%suggesting transformation of our algorithm description for site
%percolation to for bond percolation, and helping improve the
%readability of this algorithm. 
W.G.Y. was supported by the China
Postdoctoral Science Foundation and the Shanghai Postdoctoral
Science Foundation, and is supported by CUHK 4288/00P 2160148.
R.T. was supported by the National Natural
Science Foundation of China (No. 19774020), the foundation of
Educational Ministry of China (No. 99024610), and Shanghai
Research Center of Applied Physics.

%\vskip 1cm

\begin{references}
\bibitem{site} Percolation, of course, can be defined for sites. In this case,
each site is taken to be occupied with probability $p$ or vacant
with probability $1-p$, and a bond is open if and only if both its
end sites are occupied. But in the language of graph theory bond
percolation is more general because the graphs created by site
percolation are induced subgraphs of the lattice.
\bibitem{note} It is not true for
$d=2$ percolation with periodic lateral boundary conditions.
\bibitem{data} Configurations are produced by randomly
producing the bond status array: $1\equiv{\rm open}$ and
$0\equiv{\rm closed}$. Open bonds appear with an independent
probability $p$.
\end{references}
\end{document}